\documentclass[12pt,article]{article}
\usepackage{amsmath,amssymb,amsthm,bbm,graphicx,psfrag}
\usepackage[english]{babel}

\newcommand{\R}{\mathbb{R}}
\newcommand{\Z}{\mathbb{Z}}
\newcommand{\A}{\mathcal{A}}

\newtheorem{theorem}{Theorem}[section]

\title{Stochastic Growth in One Dimension and Gaussian Multi-Matrix Models}
\author{Patrik L. Ferrari, Michael Pr\"{a}hofer, Herbert Spohn\\
Zentrum Mathematik and Physik Department,\\
Technische Universit\"{a}t M\"{u}nchen, D-85747 Garching \\
ferrari@ma.tum.de, praehofer@ma.tum.de, spohn@ma.tum.de}

\begin{document}

\maketitle

\begin{abstract}
We discuss the space-time determinantal random field
which arises for the PNG model in one dimension and resembles the one for
Dyson's Brownian motion. The information of interest for growth
processes is carried by the edge statistics of the random field and
therefore their universal scaling is related to the edge properties
of Gaussian multi-matrix models.
\end{abstract}

\section{Introduction}

There is a huge variety of stochastic growth processes and we
refer to the recent monographs by Barab\'{a}si, Stanley\cite{BS}
and Meakin\cite{Mk}. Amongst them the
KPZ (Kardar, Parisi, Zhang) growth models enjoy a particular
popularity in theoretical circles. Prototypes are the Eden growth,
where each perimeter site of the current cluster is filled after
an exponentially distributed waiting time, and ballistic
deposition, where there is a low intensity random flux of incoming
particles which then attach to the current surface profile. Thus
in broad terms KPZ growth is characterised by being stochastic
with a
\begin{itemize}
\item local growth rule,
\item smoothening mechanism.
\end{itemize}
The list of experiments well described in terms of KPZ is rather
short, see Mylls {\it et al.}\cite{MMM} for recent experiments on slow combustion
fronts. But one obvious
reason for its popularity is that KPZ growth is
rather close to the stochastic dynamics commonly studied in
Statistical Mechanics with the fine twist that it does not satisfy
detailed balance (KPZ growth is not stochastically reversible).
There is a second reason, however: through a Cole-Hopf type
transformation KPZ growth maps to directed first passage
percolation. Thereby techniques and insights from the theory of
disordered systems come into play. In fact, there are some rather
close analogies, one example being the rigorous discussion of the
overlap for directed polymers\cite{CSY}.

Just before the previous ICMP congress it came as a total surprise
that KPZ growth in one spatial dimension (the surface is the graph
of a function over ${\R}$) is linked to Gaussian random matrices.
For example, the height, $h(\tau)$, at time $\tau$ above a given
reference point grows at constant speed $v_0$ with some random
fluctuations, \begin{equation} \label{1.1} h(\tau) = v_0 \tau + \tau^{1/3} \xi
\end{equation} for large $\tau$. If at the reference point the macroscopic profile has a non-zero curvature, then $\xi$ has the same distribution function as the
largest eigenvalue of a GUE random matrix, which in the
community is known as the Tracy-Widom distribution function $F_2$,
$\mathbb{P}(\xi \leq x)=F_2 (x)$. $F_2$ is related to
the Hastings-McLeod solution of the Painlev\'{e} II
differential equation. The scaling exponent $1/3$ and the random amplitude
$\xi$ in (\ref{1.1}) are expected to be valid for all one-dimensional KPZ growth
models.

The purpose of our contribution is to explain how this connection
arises. The rough indication can be given already now: For some
very particular growth models in the KPZ class, there is a
determinantal process in the background. Its structure has some
similarity to the determinantal process appearing in the context
of Gaussian
multi-matrix models. Since the relevant information is linked to edge
scaling, in the scaling limit the GUE edge distribution arises.
\section{GUE, Dyson's Brownian motion, and the Airy process}

Dyson\cite{D} considers an Ornstein-Uhlenbeck process,
$A(t)$, $t \in \R$, on the space of $N \times N$ Hermitian matrices.
Its transition probability is given through a Mehler type formula as
\begin{equation} \mathbb{P}(A(t) \in dA'|A(0) = A) = \frac{1}{Z}\exp [ - \frac{1}{N}
\text{Tr}(A' - qA)^2/(1-q^2)] dA'
\end{equation}
with $q = e^{-t}$. In particular, for $t \to \infty$ the
process converges to the GUE ensemble given by
\begin{equation} Z^{-1} \exp [- \frac {1}{N} \mathrm{Tr} A^2]dA\,.
\end{equation}
For the joint distribution of the stationary process
at times ordered as $t_0 < t_1 < ... < t_m$
one obtains therefore
\begin{eqnarray} \label{a}
&&\mathbb{P}(\{A(t_j) \in dA_j, j =0,1,...,m\})\nonumber\\
&&= \frac{1}{Z}\exp \Big[ - \frac{1}{N}\Big(\sum_{j=1}^m
\text{Tr}(A_j - q_j A_{j-1})^2/(1-q_j^2) +\text{Tr}A_0^2\Big)\Big]
\prod_{j=0}^m dA_j
 \end{eqnarray}
with $q_j = \exp[-(t_j - t_{j-1})]$. (\ref{a}) is the Gaussian
multi-matrix model of the title.

The process $A(t)$ induces a
process on the eigenvalues $\lambda_N (t) \leq ...\leq
\lambda_1(t)$ of $A(t)$. It is stationary by construction and
happens to be again Markov. In fact, it satisfies the set of stochastic
differential equations
\begin{equation} \label{1.2}
d \lambda_j (t) =
\Big(-\frac{1}{N}\lambda_j(t) + (\beta/2) \sum^N_{i=1,i\not= j}
\big(\lambda_j(t)- \lambda_i(t)\big)^{-1}\Big)dt + db_j(t)\,, \quad
j=1,...,N \,,
\end{equation}
with $\{b_j(t),\, j=1,...,N\}$ a collection of $N$ independent
standard Brownian motions. In case of Hermitian random
matrices $\beta = 2$.
The repulsive drift ensures that
$\lambda_{j+1}(t)<\lambda_j(t)$ for all $t$, any $j$. Clearly
(\ref{1.2}) has the unique stationary distribution
\begin{equation} \frac{1}{Z}\exp\Big[-\frac{1}{N}
\sum^N_{j=1}\lambda^2_j\Big] \prod_{1\leq i<j \leq N}
|\lambda_i - \lambda_j|^{\beta} \prod^N_{j=1} d\lambda_j\,.
\end{equation}

There is a way to rewrite (\ref{1.2})
which turns out to be computationally very powerful.
We define a random field $\phi_N$ over $\R^2$ through
\begin{equation} \label{1.3} \phi_N(x,t) = \sum^N_{j=1} \delta(x-\lambda_j (t))\,.
\end{equation}
Then $\phi_N$ is determinantal in the sense that its moments
(correlation functions) have a determinantal structure of the form
\begin{equation} \label{1.4}
\langle \prod^m_{j=1} \phi_N (x_j,t_j)\rangle =
\det\{R_N(x_i,t_i;x_j,t_j)\}^m_{i,j=1}
\end{equation}
for distinct times $t_1,\ldots,t_m$. The defining kernel $R_N$ can be written in terms of the harmonic oscillator Hamiltonian
\begin{equation} \label{1.5} H_N
=-\frac{1}{2} \partial^2_x + \frac{1}{2N^2}x^2\,,
\end{equation}
which has eigenvalues $E_n = n/N$, $n=1,2,...$ . Let $K_N$ be the
Hermite kernel which is the spectral projection onto $\{H_N \leq 1\}$.
Then
\begin{equation} \label{1.6}
R_N(x,t;x',t') = \Big(e^{-tH_N} \big(K_N -
\text{1\hspace{-2pt}l}\Theta(t-t')\big)e^{t'H_N}\Big)(x,x')
\end{equation}
with $\Theta(t)=1$ for
$t>0$ and $\Theta(t)=0$ for $t \leq 0$. In particular, at $t=0$
(or any other time by stationarity)
\begin{equation} \label{1.7}\langle
\prod^m_{j=1} \phi (x_j,0)\rangle =
\det\{K_N(x_i,x_j)\}^m_{i,j=1}\,,
\end{equation}
as well known from the random matrix bible by Mehta\cite{Mt}.

A natural question is to study the statistics of lines close to
the, say upper, edge. In our units the top line fluctuates at
level $\sqrt{2}N$, as can be seen by equating the Fermi
energy $E_{\textrm{F}}$,
$E_{\textrm{F}}=1$, with the energy of the confining potential of
(\ref{1.5}). Thus we shift our attention to $x=\sqrt{2}N$ and
linearise there the potential, a procedure which should be
accurate for large $N$. Then the imaginary time Schr\"{o}dinger
equation for (\ref{1.5}) goes over to
\begin{equation} \label{1.8}
\partial_t
\psi = (-\frac{1}{2} \partial^2_x + \frac{1}{N} \sqrt{2}x)\psi\,.
\end{equation}
It becomes $N$-independent under the scaling $t \leadsto
N^{2/3}t$, $x \leadsto N^{1/3} x /\sqrt{2}$, resulting in
the Schr\"{o}dinger equation with Airy Hamiltonian
\begin{equation} \label{1.9}
\partial_t \psi = H \psi\,, \quad H=
-\partial_x^2 + x \,.
\end{equation}

We have identified the edge scaling and
conclude that, in the sense of convergence of moments,
\begin{equation}\label{1.10}
\lim_{N \to \infty} \frac{1}{\sqrt2} N^{1/3}
\phi_N(\sqrt2 N + \frac{1}{\sqrt2} N ^{1/3} x \,,\; N^{2/3} t) =
\phi (x,t)\,. \end{equation} The prefactor comes from the spatial volume
element when integrating both sides in (\ref{1.10}) over a
compactly supported test function. $\phi$ is called the Airy
random field. Since $\phi_N$ is determinantal, so must be its
limit. Hence, for distinct times $t_1,...,t_m$,
\begin{equation}\label{1.11}
\langle \prod^m_{j=1} \phi (x_j,t_j)\rangle =
\det\{R(x_i,t_i;x_j,t_j)\}^m_{i,j=1}\,,
\end{equation}
compare with (\ref{1.4}). The defining kernel is now given through
\begin{eqnarray}\label{1.12}
R(x,t;x',t') &=& \Big(e^{-tH}
\big(K-\text{1\hspace{-2pt}l}\Theta(t-t')\big)e^{t'H}\Big)(x,x') \\
&=& \text{sign}(t'-t) \int d\lambda
\Theta(\lambda(t-t'))e^{\lambda(t'-t)} \text{Ai}(x-\lambda)
\text{Ai}(x'-\lambda)\,,\nonumber
\end{eqnarray}
where sign$(t)=1$ for $t \geq 0$ and sign$(t)= -1$ for $t<0$, Ai
the Airy function, and $K$ the spectral projection onto $\{H \leq
0\}$. $K$ is known as Airy kernel. For fixed $t$ the limit is
studied by Forrester\cite{F}, and Tracy, Widom\cite{TW1}. Some
aspects of the multi-matrix model are discussed by Eynard\cite{E}.

The top line of the Airy random field is called the Airy process\cite{PS1},
denoted by $\A(t)$.
Its joint distributions can be written in a concise way. Let
$t_1<...<t_m$ and let us define the operator, $R^{(m)}$, on
$L^2(\R \times \{1,...,m\},dx)$ through the integral kernel
$R(x,t_i;x',t_j)\chi(\{x>\xi_i\})\chi(\{x'>\xi_j\})$,
$i,j=1,...,m$. Then $R^{(m)}$ is trace class and
\begin{eqnarray}\nonumber
\mathbb{P}(\{\A(t_1)\leq \xi_1,...,\A(t_m) \leq
\xi_m\})&=& \mathbb{P}(\{\phi(x,t_j)=0\quad\text{for}\quad x> \xi_j\,,\, j=1,...,m\})\\
&=& \det (1-R^{(m)})\,.\label{1.13}
\end{eqnarray}
The Airy process has continuous sample paths and is stationary, by
construction. Some more explicit expressions for joint
distributions are given in \cite{AvM,TW2,TW3,W}. In particular,
$\langle \A(0)^2 \rangle = 0.81325...$, $\langle (\A(0)-\A(t))^2
\rangle = 2|t|$ for small $t$, and
$\langle
\A(0)\A(t)\rangle - \langle \A(0)\rangle^2 = t^{-2} +
\mathcal{O}(t^{-4})$ for large $t$.
\section{Polynuclear growth, its determinantal random field, and edge scaling}

For the polynuclear growth (PNG) model the height function at time
$\tau$ takes integer values, $h(x,\tau) \in \Z, \, x\in \R, \,
\tau \geq 0$. $x \mapsto h(x,\tau)$ has jumps of unit size only.
Upward steps move, in the $x$ direction, with velocity $-1$ and downward steps with
velocity 1. They annihilate each other upon collision, which is
the required smoothening mechanism. The surface grows through
nucleation. At such an event, say $(x_1, \tau_1)$, the height
$h(x,\tau_1)$ is increased by one unit at $x_1$, thereby creating
an upward and downward step, which move apart immediately under
the deterministic part of the dynamics. The nucleation events are Poisson in space-time with intensity 2. We consider the droplet
geometry which is enforced by allowing nucleations only in the
space interval $[-\tau, \tau]$. Initially $h(x,0)=0$ and, by
assumption, $h(x,\tau)=0$ for $|x|>\tau$. Obviously, according to
our criteria of the Introduction, the PNG model is in the
KPZ class.

For PNG, as just described, there is no determinantal process in sight. The miracle happens through the RSK (Robinson, Schensted, Knuth) construction. The idea is to extend the model with additional lines which record the information lost in annihilation events. Thus we introduce the lines $h_{j}(x,\tau)$, $j=0,-1,-2,\ldots,$ and set $h_0(x,\tau)=h(x,\tau)$. Initially $h_j(x,0)=j$. $h_0$ evolves according to the PNG specified above. Given $h_0(x,\tau)$ all lower lying lines have a deterministic dynamics. The steps of line $h_j$, $j \leq -1$, move with velocity $\pm 1$, as before. A nucleation, say at space-time point $(x,\tau)$, takes place whenever in line $j+1$ an upward and downward step annihilate each other at $(x,\tau)$. Thus $x \mapsto h_j(x,\tau)$ has jumps of unit size only, $h_j(x,\tau)=j$ for $|x|>\tau$, and $h_j(x,\tau)>h_{j-1}(x,\tau)$. Furthermore there
is a random index $j_0$ such that, for $j \leq j_0$,
$h_j(x,\tau)=j$ for all $x$. Let us set, at fixed time $\tau$,
\begin{equation}
\eta_\tau (j,t) =
\begin{cases}1&
\text{if there is a height line passing through}\, (j,t)\,, \\
0 & \text{if there is no such line\,,}
\end{cases}
\end{equation}
with $j \in \Z$, $t \in [-\tau,\tau]$. $\eta_{\tau}(j,t)$ is a
determinantal random field over $\Z \times [-\tau,\tau]$. Its top
line is the object of interest, since it coincides with PNG.

To write down the defining kernel for $\eta_\tau$ we set, as
operators on $l_2=l_2(\Z)$,
\begin{eqnarray}
H_{\text{d}} \psi(j)&=& -\psi(j-1)- \psi(j+1)\,, \\
H_{\tau} \psi(j)&=& -\psi(j-1)- \psi(j+1) + \frac{j}{\tau} \psi (j)\,,
\end{eqnarray}
and $B_\tau$ the
spectral projection on $\{H_\tau \leq 0\}$. $B_\tau$ is known as
discrete Bessel kernel. Then
\begin{equation} \label{1.17}
R_\tau(j,t;j',t')=\Big(e^{-tH_{\text{d}}} \big( B_\tau -
\text{1\hspace{-2pt}l}\Theta
(t-t')\big)e^{t'H_{\text{d}}} \Big)_{jj'}\,.
\end{equation}
The moments of $\eta_\tau
(j,t)$ are given by the formula analogous to (\ref{1.4}). For
large $\tau$, $h_0(x,\tau) \cong 2 \sqrt{\tau^2-x^2}$, $|x|<
\tau$. Thus $\eta_\tau (x,t)$ is not stationary in $t$, which is
reflected in (\ref{1.17}) by the fact that $H_{\text{d}}
\not= H_\tau$.

The correct edge scaling can be guessed as for Dyson's Brownian
motion, where in spirit $\tau$ is equated with $N$. We focus our
attention on a space-time window of width $\tau^{2/3}$ and height
$\tau^{1/3}$ centred at $t=0$ and $j=2\tau$. Properly rescaled,
compare with (\ref{1.8}), $H_\tau$ becomes
\begin{equation} \label{1.18}
H_\tau
\psi(x) = \tau^{2/3} \big(-\psi(x- \tau^{-1/3}) -\psi(x+
\tau^{-1/3})+2 \psi(x) + \tau^{1/3} (x/\tau)\psi(x)\big)\,,
\end{equation}
which converges to the Airy operator $H$
as $\tau \to \infty$. This
argument overlooks that even in rescaled coordinates the lines
have still a systematic curvature. The correct limit is thus, $[\cdot
]$ denoting integer part,
\begin{equation} \label{1.19} \lim_{\tau \to \infty}
\tau^{1/3} \eta_\tau ([2\tau +\tau^{1/3}(x-t^2)],\,\tau^{2/3}t) =
\phi (x,t)
\end{equation}
with $\phi$ the Airy random field. In particular, the top line converges to the Airy process $\A(t)$.

We summarise our discussion as
\begin{theorem} \label{the.1} Let $h(x,\tau)$ be the PNG model in
the droplet geometry. Then, in the sense of weak convergence of
finite-dimensional distributions,
\begin{equation} \label{1.20} \lim_{\tau \to
\infty} \tau^{-1/3}\big(h(t \tau^{2/3}\,,\, \tau)-2\tau\big) = \A(t)-t^2\,.
\end{equation}
\end{theorem}
\begin{proof} For $t=0$ this is the celebrated result of Baik, Deift,
Johansson\cite{BDJ} on the length of the longest increasing
subsequence of a random permutation. They use orthogonal
polynomials and Riemann-Hilbert techniques in their asymptotic
analysis. Johansson\cite{J1} develops an approach through Fredholm
determinants, which is also the basis of the proof by
Pr\"{a}hofer, Spohn\cite{PS1} for the time-extended case.
\end{proof}
Theorem \ref{the.1} is stated for the reference point $x=0$. An analogous limit holds for any other reference point $x= a\tau$, $|a|<1$.

For a space-time discrete version of the PNG droplet Johansson\cite{J2} proves the analogue of Theorem \ref{the.1} including tightness. While the details are long, it is likely that Theorem \ref{the.1} can be strengthed to weak convergence of path
measures. As established by Johansson\cite{J3}, for the Aztec
diamond the border between the frozen and disordered zones is also
governed by the Airy process.

\section{Extensions}

What is so special about the droplet? From the point of view of
growth processes other initial conditions look more natural, e.g.
the initially flat surface $h(x,0)=0$. The RSK dynamics can be
implemented for any choice of initial data and set of nucleation
events. However the determinantal property of the resulting line
ensemble is fragile. 

We illustrate our point by three examples.\medskip\\
(i) {\it half droplet}. In a discrete space-time setting this problem is
studied recently by Sasamoto and Imamura\cite{SI}. The rule is to
simply suppress all nucleation events for $x<0$. In addition there
is a source at $x=0$ with rate $\gamma$. For $0\leq\gamma<1$ the
additional mass is incorporated in the bulk with no change in the
macroscopic profile. The height fluctuations at $x=0$ satisfy
(\ref{1.1}) with $F_2$
replaced by $F_4$, i.e.\ the distribution function
of the largest eigenvalue of a GSE random matrix in the limit of
large $N$ \cite{BR}. $\gamma =1$ is critical and
$F_2$ in (\ref{1.1})
is replaced by $F_1$ \cite{BR}, i.e.\ the distribution function of the
largest eigenvalue of a GOE random matrix\cite{TW96}.
For $\gamma>1$ the macroscopic profile
develops a linear portion, starting at $x=0$,
at height $(\gamma+\gamma^{-1})\tau$ and joining tangentially the profile
for $\gamma=0$. The height fluctuations are then of order
$\sqrt\tau$ and Gaussian. The RSK dynamics results in a line
ensemble, which for fixed $\tau$ has the following weight. At
$x=\tau$ the boundary condition is $h_j(\tau,\tau)=j$. The lines
have jump size one and are not allowed to cross. Under these
constraints the jump points are uniformly Lebesgue distributed. If
this construction would be extended to $x=-\tau$ with the boundary
condition $h_j(-\tau,\tau)=j$, the resulting line ensemble is the
one of the PNG droplet. For the half-droplet the lines end at
$x=0$ and there is an
extra weight from the source at the origin, which reads
\begin{equation}\label{1.21}
\exp [(\log \gamma)\sum^0_{j=-\infty}
\big(h_{2j}(0,\tau)-h_{2j-1}(0,\tau)-1 \big)]\,.
\end{equation}
For $\gamma=1$ the extra weight (\ref{1.21}) equals $1$ and the weight for the line ensemble can be written as a product of determinants. However, the correlation functions are not determinantal because of the free boundary at $x=0$. By an identity of de Bruijn, gap probabilities are square roots of determinants, thus yielding GOE edge statistics at $x=0$ in the scaling limit.
For $\gamma =0$ the lines at $x=0$ are constrained as $h_{2j}(0,\tau)- h_{2j-1}(0,\tau)=1$, which resembles Kramers degeneracy, thus GSE.
An asymptotic analysis is available only for $\gamma =0,1$ \cite{SI} with the result that $\tau^{2/3}$ close to the origin the height statistics is governed
by the largest eigenvalue of the transition ensemble governing the
crossover from GOE, resp. GSE, at $x=0$ to GUE in the bulk. For
$\gamma>1$, at $x=0$ the top line separates the distance $(\gamma
+ \gamma^{-1}-2)\tau$ from $h_{-1}(0,\tau)$ which remains located
roughly at $2\tau$, as for $\gamma \leq 1$. This explains the
Gaussian fluctuations for $\gamma >1$. \medskip\\
(ii) {\it stationary PNG}. If initially the upward steps
are Poisson distributed with density $\rho_+$ and the downward
steps with density $\rho_-$, then PNG on the whole real
line is a space-time stationary growth process, provided $\rho_+
\rho_- =1$. To determine $h(0,\tau)$ it suffices to know the
nucleation events in the backward light cone $\{(x,t)| 0\leq
t\leq \tau,\,|x|\leq \tau-t\}$. In fact it also suffices to only
know the nucleation events in the forward light cone of the
origin, $\{(x,t)| 0\leq t\leq \tau,\,|x|\leq \tau\}$. The reason
is that for stationary PNG model the intersection points of the diagonals $\{x=\pm t\}$ with the world lines for steps are again Poisson
distributed. Thus we arrive at a set-up rather similar to the PNG
droplet. The additional feature is that there are sources of
nucleation events at the boundaries $\pm \tau$. The sources are
Poisson in time with rates $\alpha_+$, resp. $\alpha_-$, $\alpha_+
\geq 0$, $\alpha_-\geq 0$, stationarity being equivalent to
$\alpha_+\alpha_- = 1$. As before, a line ensemble is generated through the RSK dynamics. At time $\tau$, the boundary conditions
$h_j (\pm \tau,\tau)=j$, $j=-1,-2,...,$ hold. However, $h_0(\pm \tau,\tau)$ can now take arbitrary positive integer values. They have geometric weight, i.e. the weight \begin{equation} \label{1.21a} \exp [(\log \alpha_{\pm}) h_0(\pm \tau,\tau)]\,. \end{equation} The step points are
Lebesgue distributed, constrained by non-crossing and the boundary
conditions at $\pm\tau$. Let $\Omega$ be the collection of all
such lines. By running RSK backwards in time, one will end up with
$h_j(0,\tau)=j$ and $h_0(0,\tau)=n$ for $\tau >0$ but sufficiently
small. $\Omega$ decomposes accordingly as $\Omega= \cup_{n\geq 0}
\Omega_n$. $\Omega_n$ has the total weight $(\alpha_+ \alpha_-)^n
Z_0$, $Z_0= \exp [t(\alpha_+ - \alpha_+^{-1}) + t(\alpha_- -
\alpha_-^{-1})]$. Stationary growth corresponds to the sector
$\Omega_0$ carrying weight $\mathbb{W}_0$ and probability
$\mathbb{P}_0 = \mathbb{W}_0 /Z_0$ for the lines. However, only
the line ensemble with weight $\mathbb{W}$ on $\Omega$ is
determinantal. Fortunately, one has the simple relationship
\begin{eqnarray} \label{1.21b}
\mathbb{P}_0(\{h_0(0,\tau)=k\})&=& Z_0^{-1} \mathbb{W}_0(\{h_0(0,\tau)=k\})\\
&=&Z_0^{-1}\big(\mathbb{W}(\{h_0(0,\tau)=k\})-\alpha_+\alpha_-
\mathbb{W}(\{h_0(0,\tau)=k-1\})\big)\,.\nonumber
\end{eqnarray}
Stationary PNG is described by the spatial derivative of a
determinantal process. (\ref{1.21b}) serves as a starting point
for an asymptotic analysis, see Baik, Rains\cite{BR2} and
Pr\"{a}hofer, Spohn\cite{PS2}.\medskip\\
(iii) {\it flat initial conditions}. One sets $h(x,0)=0$ and allows for
nucleation events on the entire real line. For the RSK
construction it is necessary to first restrict to the periodic box
$[-\ell,\ell]$. Then $h_j (-\ell,\tau)= h_j (\ell,\tau)$.
As before, the lines
are constrained on non-crossing and unit jumps. In addition they have
to satisfy
\begin{equation} \label{1.22} \min_{|x|\leq \ell} \big(h_j(x,\tau)-
h_{j-1}(x,\tau)\big) =1 \,.
\end{equation}
An asymptotic analysis of the line
ensemble with these constraints is not available. By
completely different methods Baik, Rains\cite{BR} prove that,
first taking $\ell \to \infty$, the height fluctuations at $x=0$ are
distributed as the largest GOE eigenvalue, i.e.\ $F_1$ replaces
$F_2$ in (\ref{1.1}). On this basis, a possible guess for the
full height statistics, in the limit $\tau \to \infty$, is Dyson's
Brownian motion at $\beta=1$. We will have to wait for the next
ICMP congress for a confirmation (or a counter argument).






\begin{thebibliography}{99}

\bibitem{BS} A.\ L.\ Barab\'{a}si, H.\ E.\ Stanley. {\em Fractal
Concepts in Surface Growth}, Cambridge University Press, 1995.

\bibitem{Mk} P.\ Meakin. {\em Fractals, scaling and growth far from equilibrium}, Cambridge
University Press, 1998.

\bibitem{MMM} M.\ Mylls, J.\ Maunuksela, J. \ Merikoski, J.\ Timonen,
V.\ K.\ Horvath, M.\ Ha, M.\ den Nijs. {\em Effect of columnar
defect on the shape of slow-combustion fronts},
arXiv:cond-mat/0307231.

\bibitem{CSY} F.\ Comets, T.\ Shiga, N.\ Yoshida.
{\em Probabilistic analysis of directed polymers in a random
environment: a review.} Advanced Studies in Pure Mathematics,
2003, to appear.

\bibitem{D}  F.\ J.\ Dyson. {\em A Brownian motion model for the
eigenvalues of a random matrix}, J. Math. Phys. {\bf 3},
1199--1215 (1962).

\bibitem{Mt} M.\ L.\ Mehta.
{\em Random Matrices}, Academic Press, 1990.

\bibitem{F} P.\ J.\ Forrester. {\em The spectrum edge of random matrix ensembles},
Nucl. Phys.\ {\bf B402}, 709--728 (1993).

\bibitem{TW1} C.\ A.\ Tracy, H.\ Widom. {\em Level-spacing
distributions and the Airy kernel}, Comm. Math. Phys. {\bf 159},
151--174 (1994).

\bibitem{E} B.\ Eynard. {\em Correlation functions of eigenvalues of
multi-matrix models, and the limit of a time dependent matrix},
J.\ Phys.\ A {\bf 31}, 8081--8098 (1998).

\bibitem{PS1}  M.\ Pr\"{a}hofer, H.\ Spohn.
{\em Scale invariance of the PNG droplet and the Airy process},
J.\ Stat.\ Phys.\ {\bf 108}, 1071--1106 (2002).

\bibitem{AvM} M.\ Adler, P.\ van Moerbecke. {\em
A PDE for the joint distribution of the Airy process},
arXiv:math.PR/0302329.

\bibitem{TW2} C.\ A.\ Tracy, H.\ Widom. {\em A system of
differential equations for the Airy process}, Elect. Comm. Prob.
{\bf 8}, 93--98 (2003).

\bibitem{TW3} C.\ A.\ Tracy, H.\ Widom. {\em Differential equations for Dyson processes}, arXiv:math.PR/0309082.

\bibitem{W} H.\ Widom. {\em On asymptotics for the Airy process},
arXiv:math.PR/0308157.

\bibitem{BDJ}  J.\ Baik, P.\ Deift, K.\ Johansson.
{\em On the distribution of the length of the longest increasing
subsequence in a random permutation}, J. Amer. Soc. {\bf 12},
1189--1178 (1999).

\bibitem{J1}  K.\ Johansson. {\em Non-intersecting paths, random tilings and
random matrices}, Probab.\ Theory Relat.\ Fields {\bf 123}, 225--280
(2002).

\bibitem{J2}  K.\ Johansson. {\em Discrete polynuclear growth and determinantal
processes},\hspace{3cm}\phantom{0} arXiv:math.PR/0206208.

\bibitem{J3}  K.\ Johansson. {\em The arctic circle boundary and the Airy process},
arXiv:math.PR/0306216.

\bibitem{SI} T.\ Sasamoto, T.\ Imamura. {\em
Fluctuations of a one-dimensional polynuclear growth model in a
half space}, arXiv:cond-mat/0307011.

\bibitem{BR} J.\ Baik, E.\ Rains.
{\em Symmetrized random perturbations. In: Random Matrix Models
and Their Applications}, P. Bleher, A. Its, eds., MSRI
Publications {\bf 40}, 1--19, Cambridge University Press (2001).

\bibitem{TW96} C.\ A.\ Tracy, H.\ Widom. {\em On orthogonal and symplectic matrix ensembles}, Commun. Math. Phys. {\bf 177}, 727--754 (1996).

\bibitem{BR2} J.\ Baik, E.\ Rains.
{\em Limiting distributions for a polynuclear growth model with
external sources}, J. Stat. Phys. {\bf 100}, 523--541, (2000).

\bibitem{PS2}  M.\ Pr\"{a}hofer, H.\ Spohn.
{\em Exact scaling functions for one-dimensional stationary KPZ
growth}, arXiv:cond-mat/0212519, J.\ Stat.\ Phys., to appear.

\end{thebibliography}
\end{document}